\documentclass[preprint]{aastex}
\usepackage{graphicx}
\usepackage{natbib}
\usepackage{color}

\shorttitle{Sun: radio radiation - Solar cycle - Sun: radius}
\shortauthors{Selhorst et al.}

\begin{document}

\title{Planetary Transits with the ALMA Radio Interferometer}

\author{C. L. Selhorst }
\affil{IP\&D - Universidade do Vale do Para\'iba - UNIVAP, S\~ao Jos\'e dos Campos, SP, Brazil }
\email{caius@univap.br}
\and
\author{C. L. Barbosa }
\affil{IP\&D - Universidade do Vale do Para\'iba - UNIVAP, S\~ao Jos\'e dos Campos, SP, Brazil }
\and
\author{Adriana V\'alio}
\affil{CRAAM, Universidade Presbiteriana Mackenzie, S\~ao Paulo, SP, Brazil}

\begin{abstract}
{Planetary transits are commonly observed at visible wavelengths. Here we investigate the shape of a planetary transit observed at radio wavelengths. Solar maps of the Sun at 17 GHz are used as a proxy for the stellar eclipse by several sizes of planets from Super-Earths to Hot Jupiters. The relative depth at mid transit is the same as observed at visible wavelengths, but the limb brightening of the stellar disk at 17 GHz is clearly seen in the shape of the transit light curve. Moreover, when the planet occults an active region the depth of the transit decreases even further, depending on the brightness of the active region relative to the surrounding disk. For intense active region, with 50 times the brightness temperature of the surrounding disk, the decrease can supersede the unperturbed transit depth depending on the size of the eclipsing planet. For a Super-Earth ($R_p=0.02 R_s$) crossing, the decrease in intensity is 0.04\%, increasing to 0.86\% in the case when a strong active region is present. On the other hand, for a hot Jupiter with $R_p=0.17 R_s$, the unperturbed transit depth is 3\% increasing to 4.7\% when covering this strong active region. This kind of behavior can be verified with observation of planetary transits with the ALMA radio interferometer.
} 
\end{abstract}
 

\maketitle

\section{Introduction}

Exoplanets are a common feature in the Galaxy and perhaps in the Universe. Since the discovery of the first exoplanets, more than a decade ago, the number of candidates and confirmed exoplanets is growing, as the techniques to detect them are constantly being improved. More than 900 planets were confirmed so far \cite[http://exoplanet.eu]{Schneider2011}, and more than two thousand candidates are waiting for confirmation \citep{Wright2011}. Exoplanets have been identified in different environments, such as orbiting pulsars \citep{Wolszczan1992}, around young low mass stars \citep{Riaud2006} or even orbiting stars as massive as Formalhaut \citep{Kalas2005}.

Five techniques can be used to discover exoplanets: measuring the gravitational pull of the parent star by the planet(s) in the orbit(s) through precise radial velocities of photospheric stellar lines \cite[e.g., ][]{Campbell1988}; measuring the dimming of the light from the parent star during the transit of the planet(s) \cite[e.g., ][]{Alonso2004}; gravitational microlensing, when a suitable alignment between the planet and a background star occurs \cite[e.g., ][]{Konacki2003}; deviations from the precise timing of pulsars \cite[e.g., ][]{Wolszczan1992}; and direct imaging \cite[e.g., ][]{Kalas2005}. While the method of the radial velocities can give us the lower limit for the planet mass, the method of planetary transits can gives us the radius of the planet. The improvement of the observational techniques has led to spectrographs  with resolution as high as 100,000, and space born telescopes that push the detection limits to Earth-mass planets \citep{Dumusque2012}, or planets as small as the Moon \citep{Barclay2013} orbiting sun-like stars.

Planetary transits have been used not only to detect, but also to characterize exoplanets. Even backyard telescopes can be used to find new candidates, requiring only moderate periods of continuous monitoring. Until now, only optical surveillance was successful on detecting exoplanets, although recently, the transit of the well known exoplanet HD 189733b was observed in X-rays \citep{Poppe2013}. Additionally, attempts to observe exoplanets at longer wavelengths, such as radio, were restricted to the detection of low frequency emission from Hot Jupiters,  like  Jupiter itself \citep{Hallinan2013}. So far, no detection has been reported.

Aiming to study spots in the surface of solar type stars, \cite{Adriana2003} proposed that planetary transits could be used to  detect the presence of spots, as well as, some of their physical parameters, such as size, temperature, and intensity. The method simulates the decrease in stellar flux caused by the planet, when it crosses the star surface. Solar white light images were used to simulate the star, whereas the planet was considered an opaque disk. Because starspots are cooler than the surrounding photosphere, when the planet cross over the spot, a small increase in the light curve should be observed.

The method above was applied to 77 consecutive transit light curves of a G7 star (CoRot-2) with $0.902R_\odot$, which is orbited by a hot Jupiter planet with $0.172R_s$. The results suggested that the spots on CoRoT-2 are larger and cooler than sunspots \citep{Adriana2010}.

Whereas the presence of strong magnetic field concentrations at solar surface prevent the overturn of convective plasma cells, resulting in photospheric dark spots, these same magnetic field lines in the atmosphere above sunspots trap the plasma within the loops. This cause the local density to increase
and consequently enhance the bremsstrahlung emission, making this same region to appear brighter than the surrounding areas at radio frequencies. Moreover, depending of the magnetic field intensity the { gyro-frequency} emission could also become important. 

In this work, we will { discuss}  the use of planetary transits to characterize the radio emission coming from the parent star, as well as the physical processes involved.  Such kind of observations should become possible in the next few years with the operation of the large radio interferometer, the Atacama Large Millimeter/submillimeter Array (ALMA).
  
\section{Simulations}

To estimate the stellar radio flux, we used solar maps at 17~GHz, obtained by the Nobeyama Radioheliograph \cite[NoRH, ][]{Nakajima1994}. The 17~GHz maps have { 10'' spatial resolution}, and the median brightness temperature is normalized to $10^4$~K. Since the main interest of this work is in the range of radio frequency that will be observed by ALMA in its final form (31.3 to 950~GHz), the planet can be simulated as opaque circular disk, as proposed by \cite{Adriana2003} for white light simulations. The planet radius, $R_p$, is expressed as a fraction of the stellar radius $R_s$. During the planetary eclipse, the total intensity is obtained, every 2 minutes, from the sum of all pixel values, which yields the light curve intensity at that instant. 

\begin{figure}[!h]
\centerline{ {\includegraphics[width=14.cm]{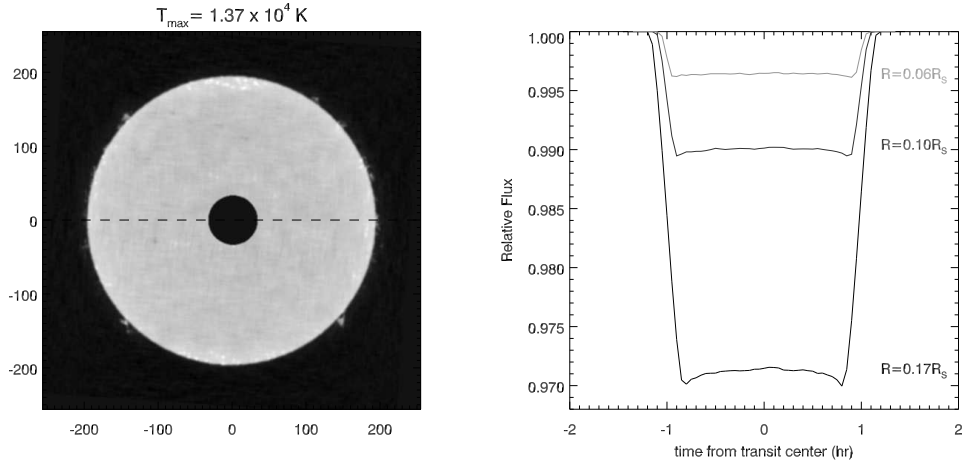}}}
\caption{a) Solar map without active regions with the opaque disk of the planet. The dashed line represents the path of the planet. b) Light curves for transits of planets with distinct sizes: $0.06, 0.10,$ and $0.17~ R_s$.}
\label{fig:qs}
\end{figure}

First, we estimate the effect of  the planetary transit in the emission of a quiet star, i.e.,  without spots or active regions. The 17 GHz solar map used for this purpose was taken on July 18th, 2009, and presented a maximum brightness temperature of $T_B=1.37\times 10^4$~K. After this map choice, different size planets were simulated to cross the stellar equator. Figure~\ref{fig:qs}a shows the stellar radio disk and the opaque disk of a Jupiter size planet ($R_p = 0.1 R_s$). The changes in the stellar light curve due to the planetary transits are plotted in Figure~\ref{fig:qs}b, whereas the intensity reduction for the largest planet ($R_p=0.17R_s$) was  $3.00\%$, the changes due to a Super Earth planet ($R_p=0.02R_s$) were smaller than $0.05\%$. In Table 1, we summarized the intensity reduction due to the transit of different size planets. 

One difference from observations at visible wavelengths is that instead of the photospheric limb darkening, we observe limb brightening because radio emission at 17 GHz originates from the chromosphere where there is a positive temperature gradient. Because of this limb brightening, which characterizes the NoRH solar maps at 17~GHz \cite[see for example, ][]{Selhorst2011}, in the quiet star simulations the maximum intensity reduction occurs in the limb region. For the simulation with the largest planet, the reduction in the limb was $3.00\%$, whereas this value drops to $2.87\%$ when the planet blocks the emission coming  from center of the stellar disk.

\begin{figure}[!h]
\centerline{ {\includegraphics[width=14.cm]{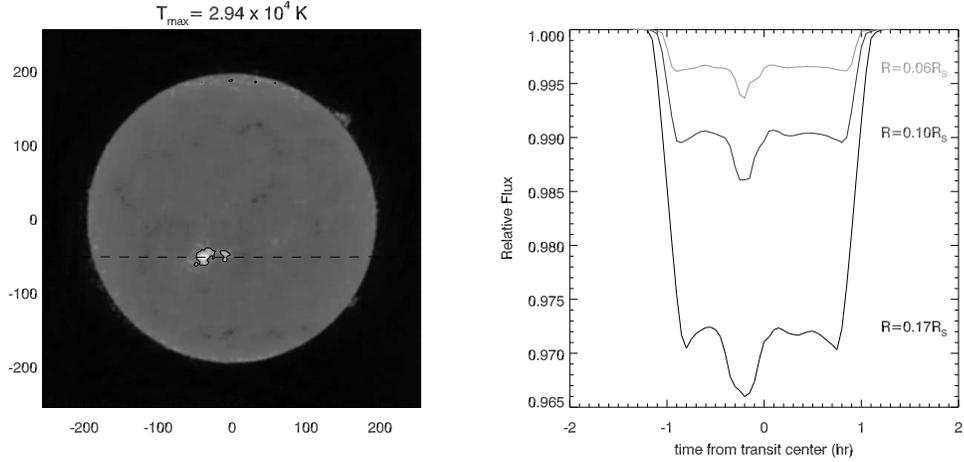}}}
\caption{a) Solar map with an active region of brightness temperature contour of $1.5 \times 10^4$~K. The dashed line represents the path of the planet, that coincides with maximum active region temperature position. b) Light curves for transits of planets with distinct sizes: $0.06,0.10$, and $0.17~ R_s$.}
\label{fig:ar1}
\end{figure}

In the second approach, we considered the presence of active regions. In this analysis the planetary orbits coincide with the position of the active region maximum temperature. The simulation shown in Figure~\ref{fig:ar1} uses the NoRH solar map observed on August 2nd, 1996, which presents a compact active region (NOAA 07981, hereafter referred to as AR-1), with maximum brightness temperature of $T_B=2.94 \times 10^4$~K and a radius of $\sim0.06R_s$ calculated at half the maximum value. The active region was delimited by a contour curve of $T_B=1.50 \times 10^4$~K. 

As can be seen in the light curves plotted in Figure~\ref{fig:ar1}b, the relative flux reduces when the planet crosses over  the active region. This behavior is contrary to what is seen at visible wavelengths when the planet crosses over a starspot, there is a small increase within the transit light curve.
These simulation results,  also listed in the Table~1, showed that for a small planet of $R_p=0.02R_s$, the maximum reduction was $0.09\%$, coinciding with the active region blockage, a difference of $0.05$ in comparison with the quiet star simulation. For the largest planet ($R_p=0.17R_s$) the flux reduction was $3.40\%$, also larger in comparison to the quiet star simulation for this planet.

\begin{figure}[!h]
\centerline{ {\includegraphics[width=14.cm]{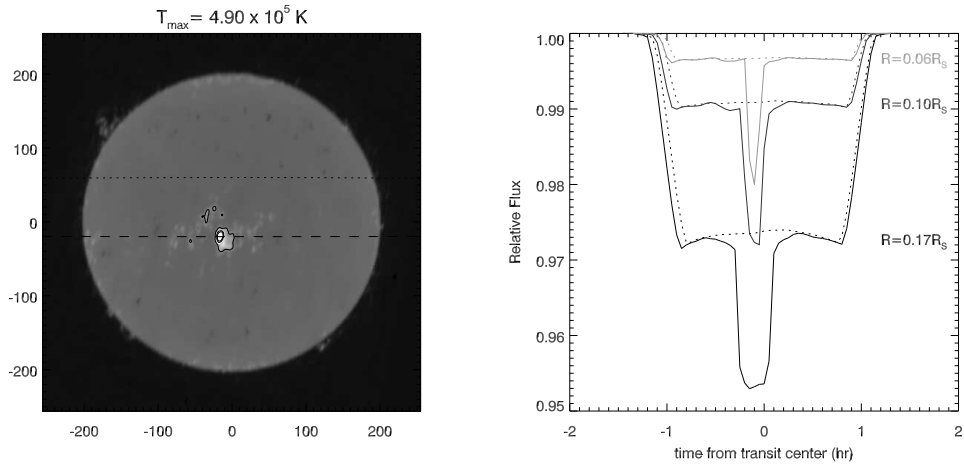}}}
\caption{a) Solar map with an intense active region with brightness temperature contours of 1.5, 5.0 and $10.0\times 10^4$~K. The dashed line represents the path of the planet, that coincides with the maximum temperature position. b) Light curves for transits of planets with distinct sizes: $0.06,0.10$, and $0.17~ R_s$. The dotted lines represent the unperturbed transits as in Figure~\ref{fig:qs}.}
\label{fig:ar2}
\end{figure}

Last we simulate planetary transits over a strong active region. For this we chose the solar map observed on December 11th, 2006 with an intense active region (NOAA 10930) of maximum brightness temperature $T_B=4.90 \times 10^5$~K (hereafter referred to as AR-2). In Figure~\ref{fig:ar2}a, the AR-2 contours delimited regions with brightness temperatures greater than 1.5, 5.0 and $10.0 \times 10^4$~K.  For this map, the simulations, plotted in Figure~\ref{fig:ar2}b, showed even greater intensity reductions, up to $4.70\%$ for planet of $R_p=0.17R_s$. Even for a Neptune size planet ($R_p=0.06R_s$) the reduction reaches $2.00\%$. It is necessary to take into account that the higher temperature intensities are confined within a region of only $\sim0.02R_s$, whereas for smaller temperatures, for example, at $T_B=1.50\times 10^4$~K  the active region radius increases to $\sim0.08R_s$.  

\begin{table}
\begin{tabular}{cccc}
\hline
\hline
Planetary size & \multicolumn{3}{c}{Maximum Intensity Reduction }\\
\hline
& Quiet Star & AR-1 & AR-2\\
\hline
0.02$R_s$ & $0.04\%$ & $0.09\%$ & $0.86\%$  \\
0.04$R_s$ & $0.18\%$ & $0.31\%$ & $1.65\%$  \\
0.06$R_s$ & $0.39\%$ & $0.63\%$ & $2.00\%$  \\
0.08$R_s$ & $0.69\%$ & $0.99\%$ & $2.36\%$  \\
0.10$R_s$ & $1.05\%$ & $1.39\%$ & $2.80\%$   \\
0.15$R_s$ & $2.35\%$ & $2.76\%$ & $4.05\%$    \\
0.17$R_s$ & $3.00\%$ & $3.40\%$ & $4.70\%$   \\
\hline
\end{tabular} 
\caption{The stellar bright intensity reduction due to the transit of different size planets. We tested the eclipses of a quiet star without active regions, and a star with an unpolarized active region, and another with a polarized one. }
\label{tab1}
\end{table}

\section{Discussion and conclusions}

In the simulations above we have used the NoRH 17~GHz radio maps to simulate the stellar emission and estimate the intensity decrease caused by the transit of an opaque planet. Considering the continuum emission of the star, the decrease caused by the planetary transit will be approximately the same for all ALMA frequencies (31.3 to 950~GHz), because the emission of the planet is basically negligible in comparison with that of the star. The main mechanism for the quiet Sun emission at 17 GHz is thermal bremsstrahlung.

\begin{figure}[!h]
\centerline{ {\includegraphics[width=14.cm]{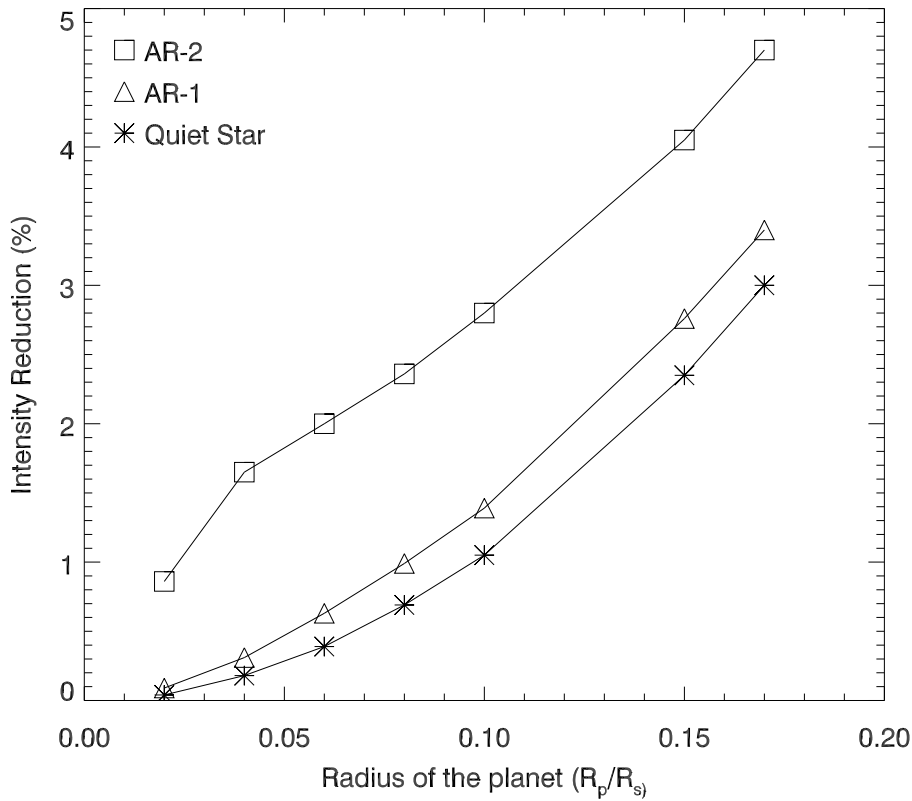}}}
\caption{Maximum intensity reduction during the transit of planets with different sizes, from 0.02 to 0.17 $R_s$. Three cases are considered: no active regions present (asterisks); moderate active region AR-1  (triangles), and strong active region AR-2 (squares).}
\label{fig:tab}
\end{figure}

We studied three cases, a star without active regions, an active region of moderate intensity (AR-1) and a strong active region (AR-2) in the path of a planet. The relative maximum decrease in the light curve during the transit is listed on Table~\ref{tab1} for different planet sizes, from Super-Earth ($0.02 R_s$), to Neptune ($0.06 R_s$), Jupiter $0.10 R_s$), and beyond. These same results are plotted in Figure~\ref{fig:tab}.

For a star without active regions (asterisks in Figure~\ref{fig:tab}), the simulations showed a maximum emission reduction of $3.00\%$ for a hot Jupiter with radius of  $0.17~R_s$ (see Figure~\ref{fig:qs}). As can be seen from Figure~\ref{fig:tab}, the intensity reduction increases as the square of the planetary radius, or planet area, as expected. The break observed in the intensity reduction of AR-2 below 0.05 $R_p=0.05 R_s$ occurs because the smallest simulated planet is smaller than the region where the gyro emission is important.

Also evident in these simulations is the presence of limb brightening, that is { predicted} by theoretical models for solar observations { at high radio frequencies ($\gtrsim15$~GHz) due to the positive temperature gradient of the solar chromosphere \cite[see: ][and references therein]{Selhorst2005a}}. This phenomenon is an intrinsic characteristic of NoRH daily 17~GHz maps. Several articles in the literature report observations of limb brightening at { high} frequencies from 33 to 860~GHz \cite[see a compilation of these results in ][]{Selhorst2003}. { Moreover, this increase of temperature at the limb is also observed at lower radio frequencies due to the optically thin solar corona \citep{Saint2012}.} Observations of limb brightening at multiple frequencies could help to model the temperature variations in the stellar atmosphere.

Whereas the main emission of the 17 GHz disk is thermal bremsstrahlung, if the magnetic field above  active regions is strong enough, the gyro-resonance emission may became important. In solar active regions, the gyro-resonance observed at 17~GHz originates mainly in the third harmonic ($\sim 2000$~G), which requires a magnetic field intensity of at least 2200~G at the photospheric level \citep{Vourlidas2006}. Moreover, due the free-free absorption, the gyro-resonance needs to occur in a region of low density. In the case of the solar atmosphere the gyro-emission will only be observed at 17GHz if it originates in the transition region or in the corona above \citep{Shibasaki1994,Shibasaki2011}.  

The active region AR-1 shown in Figure~\ref{fig:ar1}a presented a maximum brightness temperature of $T_B=2.94\times 10^4$~K, which is almost three times the brightness temperature of the quiet Sun, and a common value for 17~GHz active regions in which the bremsstrahlung is the main emission mechanism. Similar brightness temperature values are also observed at 34~GHz maps and successfully modeled as free-free radiation in the model proposed by \cite{Selhorst2008}. Unfortunately, these excess brightness temperature values observed in active region reduces drastically at higher frequencies, and do not exceed 20\% at 212  and 405~GHz \citep{Adriana2005}. If the active region (AR-1) was saturated at 20\% above the quiet star level, the decrease in the emission caused by the simulated transit of the biggest planet  would be $3.10\%$. The quadratic dependence of brightness decrease on planetary radius, or area, is also seen in Figure~\ref{fig:tab} (triangles), slightly larger than the decrease for the quiet star case, due to the small contribution of the active region to the total flux of the star.

On the other hand, a strong active region like the one depicted in the third simulation (AR-2) causes a further decrease in brightness, sometimes larger than the unperturbed transit itself, specially for the smaller planets (see Figure~\ref{fig:ar2}). The dependence of the intensity reduction on planetary radius (squares on Figure~\ref{fig:tab}) is no longer quadratic, probably indicating that the total active region flux blockage dominates over the disk area occulted by the planet.

In the model proposed by \cite{Selhorst2008}, the authors concluded that the 17~GHz gyro-resonance emission is due to the third harmonic, and the emission core is located above the transition region. However, at NoRH 34~GHz the maximum brightness temperature is completely modeled by bremsstrahlung emission.  \cite{Selhorst2009} modeled the brightness temperature observed in active region AR-2 ($T_B=4.90\times 10^5$~K) and reported that its high values are only possible with the presence of gyro-resonance.

At 34~GHz, the gyro-resonance emission will only be important if we have the presence of more intense magnetic field above the transition region. Taking into account that at radio frequencies the gyro-frequency emission originates at  harmonics between 2 and 4 \citep{Kundu2001,Vourlidas2006},  it is necessary a magnetic field of at least $\sim3000$~G to generate some appreciable emission at 34~GHz. However, these intense magnetic fields are not common in the Sun. In a statistic study between 1917 and 2004, \cite{Livingston2006} reported that only 0.2\% of the photospheric sunspots presented magnetic fields above 4000~G. 

These strong magnetic fields are not common in the Sun, however, the scenario can be completely different for a young star like CoRot-2, which is far more active than the Sun, and presents spots larger and cooler than sunspots \citep{Adriana2010}.  

Observations with ALMA at high radio frequencies will provide information on the chromosphere temperature structure through detection of planetary transits with multiple wavelengths. Models of the stellar atmosphere such as that of  \cite{Selhorst2005a} can be compared to observations to determine the temperature stratification of the stellar atmosphere.

{ To estimate the possibility of transit observation with ALMA, we considered a star like the Sun at a distance of 10 parsecs. At 345~GHz the stellar flux, assuming $T_B=5800$~K, is approximately 0.34~mJy. To be able to detect a transit of a hot giant planet the rms has to be within 3 \% of the flux, that is, 0.011~mJy. We have estimated the integration time using the ALMA Sensitivity Calculator (http://almascience.eso.org/propo\-sing/sensitivity-calculator). This can be achieved for an integration time of 1 h, considering optimal conditions of observations with ALMA, 50 antennas of 12 m, and a bandwidth of 16~GHz.

Many of the stars that host transiting planets are known to be more active than the Sun, thus their brightness temperature will be larger yielding deeper transits, easier to detect. Moreover, the area covered by spots on these stars can reach tens of a percent (for the Sun, sunspots occupy less than 1\% of its area), which will also contribute to increasing the depth of the transit. The addition of multiple transits also improves the transit detection.

Therefore, radio observations will be useful to definitely prove that the variations detected in the visible transit light curves are indeed caused by spots, since the signature of spots at radio wavelengths is exactly the opposite. Magnetic field intensities within these regions may also be estimated from the radio flux deficit caused by the blockage of the active region emission.}  
     
\acknowledgments

We would like to thank the anonymous referee whose comments greatly improved the paper. We also would like to thank the Nobeyama Radioheliograph, which is operated by the NAOJ/Nobeyama Solar Radio Observatory. C.L.S. and C.L.B. acknowledge financial support from the S\~ao Paulo Research Foundation (FAPESP), grant number 2012/08445-9.

\end{document}